\documentclass[10pt,conference]{IEEEtran}

\usepackage{amssymb}
\usepackage{amsmath}
\usepackage{pst-all}
\usepackage{graphicx}
\usepackage{color}

\newtheorem{Thm}{Theorem}
\newtheorem{Lem}[Thm]{Lemma}

\newtheorem{Pro}[Thm]{Proposition}

\newsavebox{\fminibox}
\newlength{\fminilength}
\newenvironment{fminipage}[1][\linewidth]
   {\setlength{\fminilength}{#1}\vspace{2mm}%
    \begin{lrbox}{\fminibox}\begin{minipage}{\fminilength}}
   {\end{minipage}\end{lrbox}\noindent\fbox{\usebox{\fminibox}}\vspace{2mm}}

\definecolor{Eqcolor}{rgb}{.6,.3,.3}
\definecolor{BlockDiagramColor}{rgb}{.95,.95,0.65}
\definecolor{BlockDiagramColor2}{rgb}{0.9,1,1}
\definecolor{DiagramLineColor}{named}{BrickRed}
\definecolor{DiagramLineColor2}{rgb}{0.4,0.1,0.4}
\definecolor{DiagramLineColor3}{rgb}{0.8,0.2,0.6}
\definecolor{TitleColor}{rgb}{0.6,0,0}
\definecolor{NameColor}{named}{Mahogany}
\definecolor{EdgeColor}{rgb}{0.8,0.2,0.2}
\definecolor{EmColor}{rgb}{0.7,0.3,0.3}
\definecolor{DiagramLetterColor}{rgb}{0.3,0.2,0.0}
\definecolor{ThmColor}{named}{RedOrange}
\definecolor{BulletColor}{rgb}{0.5,0,0}
\definecolor{SectionColor}{rgb}{0.9,0.1,0.1}

\definecolor{Table1}{rgb}{1,.95,.9}
\definecolor{Table2}{rgb}{1,.92,.87}
\definecolor{Table3}{rgb}{1,.89,.84}
\definecolor{Table4}{rgb}{1,.86,.81}

\definecolor{HeaderColor}{rgb}{.6,.2,.2}

\newcommand{\polytope}{
  \psset{linewidth=0.5pt}
  \psline{->}(0,0)(0,1.1)
  \psline{->}(0,0)(1.1,0)
  \psline[linestyle=dashed](0,1)(1,1)
  \psline[linestyle=dashed](1,0)(1,1)
  \pspolygon[fillstyle=solid,fillcolor=BlockDiagramColor,
    linecolor=DiagramLineColor,linewidth=1pt]
  (0,0)(0.7,1)(1,1)(1,0.2)
  \rput(0.6,0.5){$P$}
  \rput(-0.13,-0.13){$0$}
  \rput(1,-0.13){$1$}
  \rput(-0.13,1){$1$}
}

\newcommand{\cone}{
  \psset{linewidth=0.5pt}
  \psline{->}(0,0)(0,1.6)
  \psline{->}(0,0)(1.6,0)
  \pspolygon[fillstyle=solid,fillcolor=BlockDiagramColor,
    linecolor=white,linewidth=0pt]
  (0,0)(1.5,0.3)(1.05,1.5)
  \psset{linecolor=DiagramLineColor,linewidth=1pt}
  \psline(0,0)(1.5,0.3)
  \psline(0,0)(1.05,1.5)
  \rput(0.8,0.6){$K$}
  \rput(-0.13,-0.13){$0$}
  \rput(1,-0.13){$1$}
  \rput(-0.13,1){$1$}
}

\newcommand{\fundcone}[1]{
  \begin{center}
    \psset{unit=#1}
    \begin{pspicture}(0,-0.1)(3,1.5)
      \rput(0,0){\polytope}
      \rput(2,0){\cone}
    \end{pspicture}
  \end{center}
}

\newcommand{\conegen}{
  \psset{linewidth=0.5pt}
  \psline{->}(0,0)(0,1.6)
  \psline{->}(0,0)(1.6,0)
  \pspolygon[fillstyle=solid,fillcolor=BlockDiagramColor,
    linecolor=white,linewidth=0pt]
  (0,0)(1.5,0.3)(1.05,1.5)
  \psset{linecolor=DiagramLineColor,linewidth=1pt}
  \psline{->}(0,0)(1.5,0.3)
  \psline{->}(0,0)(1.05,1.5)
  \rput(0.7,1.3){$w_1$}
  \rput(1.7,0.3){$w_2$}
  \rput(0.8,0.6){$K$}
  \rput(-0.13,-0.13){$0$}
}

\newcommand{\dcone}{
  \psset{linewidth=0.5pt}
  \pspolygon[fillstyle=solid,fillcolor=BlockDiagramColor,
    linecolor=white,linewidth=0pt]
  (0,0)(-0.3,1.5)(1.5,1.5)(1.5,-1.05)
  \psline{->}(0,0)(0,1.6)
  \psline{->}(0,0)(1.6,0)
  \psline(-0.3,1.5)(0.15,-0.75)
  \psline(1.5,-1.05)(-0.75,0.525)
  \psset{linecolor=DiagramLineColor,linewidth=1pt}
  \psline(0,0)(-0.3,1.5)
  \psline(0,0)(1.5,-1.05)
  \rput(0.5,0.5){$K^*$}
  \rput(-0.13,-0.13){$0$}
  \rput(-0.6,1.7){$z^Tw_2 = 0$}
  \rput(1.8,-1.25){$z^Tw_1 = 0$}
}

\newcommand{\dualcone}[1]{
  \begin{center}
    \psset{unit=#1}
    \begin{pspicture}(-0.3,-1.5)(6,1.9)
      \rput(0,0){\conegen}
      \rput(3,0){\dcone}
    \end{pspicture}
  \end{center}
}

\newcommand{\AWGNchannel}[1]{
  \begin{center}
    \psset{unit=#1}
    \begin{pspicture}(0,0.3)(5,2.7)
      \psset{arrowsize=2pt 3,arrowlength=1.4,arrowinset=.5}
      \psset{fillstyle=solid,fillcolor=BlockDiagramColor}
      \psframe(1,0)(3,2)
      \rput(2,1.5){$0 \rightarrow +1$}
      \rput(2,0.5){$1 \rightarrow -1$}
      \psset{fillstyle=none,linecolor=DiagramLineColor}
      \psline{->}(0,1)(1,1)
      \psline{->}(3,1)(3.9,1)
      \psline{->}(4.1,1)(5,1)
      \psline{->}(4,2)(4,1.1)
      \rput(4,1){$\oplus$}
      \rput(-0.5,1){$y_i$}
      \rput(4,2.4){$\sim N(0,\sigma^2)$}
      \rput(5.5,1){$r_i$}
    \end{pspicture}
  \end{center}
}

\newcommand{\bounddist}[1]{
  \begin{center}
    \psset{unit=#1}
    \begin{pspicture}(0,-1.15)(2,1.7)
      \rput(0,0){\dcone}
      \psdot(1,1)
      \psset{linewidth=0.5pt}
      \psline{<->}(1,1)(-0.153846,0.769231)
      \psline{<->}(1,1)(0.201342,-0.140940)
      \rput(0.4,1.2){\footnotesize $\frac{\mathbf{1}^Tw_2}{\|w_2\|}$}
      \rput(1.1,0.4){\footnotesize $\frac{\mathbf{1}^Tw_1}{\|w_1\|}$}
      \rput(1.15,1.15){$\mathbf{1}$}
    \end{pspicture}
  \end{center}
}

\newcommand{\pwhamseven}{
  \psset{labelsep=2pt}
  \psset{linewidth=0.5pt}
  \psaxes[Ox=2.8,Dy=2,Dx=0.2,tickstyle=bottom,ticksize=2pt]{->}(2.8,0)(4.2,17)
  \rput(2.8,18){number of generators}
  \rput(3.5,-3.3){pseudo-weight}
  \psset{linewidth=5pt,linecolor=blue}
  \psline(3,0)(3,13)
  \psline(3.266667,0)(3.266667,9)
  \psline(3.571429,0)(3.571429,16)
  \psline(4,0)(4,4)
}

\newcommand{\pwhamfifteen}{
  \psset{xunit=2.3cm,yunit=0.04mm}
  \psset{labelsep=2pt}
  \psset{linewidth=0.5pt}
  \psaxes[Ox=2.5,Dy=100,Dx=0.5,tickstyle=bottom,ticksize=2pt]{->}(2.5,0)(5.3,670)
  \rput(2.5,720){number of generators}
  \rput(3.8,-130){pseudo-weight}
  \psset{linewidth=0.8pt,linecolor=blue}
  \psline(3,0)(3,127)
  \psline(3.0625,0)(3.0625,36)
  \psline(3.125,0)(3.125,64)
  \psline(3.169014,0)(3.169014,36)
  \psline(3.2,0)(3.2,16)
  \psline(3.266667,0)(3.266667,138)
  \psline(3.270270,0)(3.270270,64)
  \psline(3.313725,0)(3.313725,96)
  \psline(3.457143,0)(3.457143,54)
  \psline(3.521739,0)(3.521739,300)
  \psline(3.571429,0)(3.571429,412)
  \psline(3.595745,0)(3.595745,36)
  \psline(3.6,0)(3.6,144)
  \psline(3.769231,0)(3.769231,125)
  \psline(3.857143,0)(3.857143,18)
  \psline(3.903226,0)(3.903226,192)
  \psline(4,0)(4,635)
  \psline(4.090909,0)(4.090909,24)
  \psline(4.166667,0)(4.166667,96)
  \psline(4.263158,0)(4.263158,240)
  \psline(4.454545,0)(4.454545,250)
  \psline(4.5,0)(4.5,312)
  \psline(5,0)(5,25)
}

\newcommand{\pwspectrum}{
  \begin{center}
    \psset{xunit=4cm,yunit=1.8mm}
    \begin{pspicture}(0,-3)(2,45)
      \scriptsize
      \rput(-2.45,25){\pwhamseven}
      \rput(-1.15,0){\pwhamfifteen}
    \end{pspicture}
  \end{center}
}

\begin{document}

\title{Relaxation Bounds on the Minimum Pseudo-Weight of Linear Block Codes}


\author{\authorblockN{Panu Chaichanavong}
\authorblockA{Marvell Semiconductor, Inc.\\
Sunnyvale, CA 94089, USA\\
Email: panuc@marvell.com} \and
\authorblockN{Paul H. Siegel}
\authorblockA{Center for Magnetic Recording Research\\
University of California, San Diego\\
La Jolla, CA 92093, USA\\
Email: psiegel@ucsd.edu}}

\maketitle

\begin{abstract}
Just as the Hamming weight spectrum of a linear block code sheds
light on the performance of a maximum likelihood decoder,
the pseudo-weight spectrum provides insight into the performance
of a linear programming decoder. Using properties of polyhedral
cones, we find the pseudo-weight spectrum of some short codes.
We also present two general lower bounds on
the minimum pseudo-weight. The first bound is based on the column
weight of the parity-check matrix. The second bound is computed
by solving an optimization problem. In some cases, this bound
is more tractable to compute than previously known bounds
and thus can be applied to longer codes.
\end{abstract}

\section{Introduction}
Inspired by the success of iterative message-passing decoding,
there have been numerous efforts to understand its 
behavior~\cite{Wib96,FKKR01,FKV01}. Recently, Koetter
and Vontobel~\cite{KV03} presented an analysis
of iterative decoding based on graph-covering.
This analysis explains why the notion of pseudo-codeword
arises so naturally in iterative decoding.
They also showed that the set of pseudo-codewords can be
described as a polytope,
which they called the fundamental polytope.

At the same time, Feldman~\cite{Fel03} introduced a decoding algorithm
based on linear programming (LP). This decoder was successfully
applied to low-density parity-check (LDPC) codes and many turbo-like codes.
It turns out that this decoding method 
is closely related to the analysis by Koetter and Vontobel.
In particular, the feasible region of Feldman's linear program
agrees with the fundamental polytope.

For a given channel, a pseudo-weight can be defined for each
pseudo-codeword. The pseudo-weight spectrum relates to
the performance of an LP decoder in very much the same way
as the Hamming weight spectrum does to the performance
of a maximum likelihood decoder. Thus it is of interest to
find the pseudo-weight spectrum of a code. For very short codes,
this might be achieved by employing a technique related to
the dual polyhedral cone given in~\cite{CCJP99}. Some examples
of pseudo-weight spectrum calculation 
for the additive white Gaussian noise channel
will be demonstrated in Section~\ref{Sec:pwAWGN}.
For longer codes, computing the entire pseudo-weight spectrum 
becomes intractable and we have to judge the performance of a code
from bounds on the minimum pseudo-weight. Some techniques to
compute lower bounds on minimum pseudo-weight were presented
in~\cite{VK04}. In Section~\ref{Sec:bounds}, we will discuss
two new lower bounds. One is based on the column weights of the 
parity-check matrix. The other is computed by solving an optimization
problem.

\section{Linear Programming Decoding}

Let $C$ be a binary linear code of length $n$. Such a code is a linear
subspace of $\mathbb{F}_2^n$. In this paper, we will also view
$C$ as a subset of $\mathbb{R}^n$. Suppose that a codeword $y$ is
transmitted through a binary-input memoryless channel and $r$
is the output of the channel. The log-likelihood ratio $\gamma$
is defined as
$$
\gamma_i = \ln\left(\frac{\mathrm{Pr}(r_i \; | \; y_i = 0)}
{\mathrm{Pr}(r_i \; | \; y_i = 1)}\right).
$$
Any codeword $x \in C$ that minimizes the cost $\gamma^Tx$
is a maximum-likelihood (ML) codeword~\cite{Fel03}.
Thus ML decoding is equivalent to solving the problem:
\begin{center}
\begin{fminipage}
\begin{tabbing}
minimize \rule{5mm}{0mm}\= $\gamma^{T} x$\\
subject to \> $x \in C$.
\end{tabbing}
\end{fminipage}
\end{center}
Letting $H$ be a parity-check matrix of $C$, 
the feasible set of this problem can be relaxed to a polytope
$P = \{x \in \mathbb{R}^n \; | \; Bx \leq b, \; 0 \leq x_i \leq 1\}$,
where the matrix $B$ and the vector $b$ are determined from $H$
as follows~\cite{Fel03}. For a row $h$ of $H$, let $U(h)$
be the support of $h$, i.e., the set of positions of $1$ in $h$.
Then $Bx \leq b$ consists of the following inequalities:
for each row $h$ of $H$ and for each set $V \subseteq U(h)$ such
that $|V|$ odd,
\begin{equation}
\label{eq:polytopeinequality}
\sum_{i \in V} x_i - \sum_{i \in U(h) \setminus V} x_i \leq |V|-1.
\end{equation}
Now, the problem is transformed to a linear program.
This approach, introduced by Feldman~\cite{Fel03},
is called {\em{linear programming (LP) decoding}}.
The polytope $P$ is called the {\em{fundamental polytope}} by Koetter and
Vontobel~\cite{KV03} (Fig.~\ref{Fig:polytopecone}).
It has the property that a $0$-$1$ vector is in
the polytope if and only if it is a codeword of $C$.
Thus, if a $0$-$1$ vector is a solution to the linear program,
it must be an ML codeword. However, unlike in ML decoding, the output
of the LP decoder may not be a $0$-$1$ vector, 
in which case the decoder simply 
declares an error.

\section{Error Analysis}

\begin{figure}[ht]
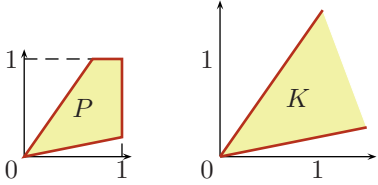

\fundcone{13mm}
\caption{Conceptual pictures of the fundamental polytope and
the fundamental cone}
\label{Fig:polytopecone}
\end{figure}

The fundamental polytope has a symmetry property that allows us to
assume without loss of generality that the all-zeros codeword
is transmitted, provided that the channel is a binary-input
output-symmetric channel~\cite{Fel03}. 
Roughly speaking, the fundamental polytope
``looks'' the same from every codeword. Therefore we assume that
the all-zeros codeword is transmitted and remove 
from the linear program all inequality
constraints that are not active at the origin.
(An inequality $f(x) \leq \alpha$ is active at a point $x^*$
if $f(x^*) = \alpha$.)
We obtain a new linear program, which we will call \textsc{LPcone}:
\begin{center}
\begin{fminipage}
\begin{tabbing}
minimize \rule{5mm}{0mm}\= $\gamma^{T} x$\\
subject to \> $x \in K = \{x \in \mathbb{R}^n \; | \; Ax \leq 0, 
\; x_i \geq 0\}$,
\end{tabbing}
\end{fminipage}
\end{center}
where $A$ is the submatrix of $B$ corresponding to zero-entries of
$b$. 
The feasible set $K$ is a polyhedral cone and 
it is called the {\em{fundamental cone}}~\cite{KV03}
(Fig.~\ref{Fig:polytopecone}).

Assuming that the zero codeword is transmitted, the probability of error
of \textsc{LPcone} is the same as that of the linear program
in the previous section. To see this, suppose that the zero vector 
is a solution to the original linear program. Then $\gamma^Tx \geq 0$
for all $x \in P$. It can be shown that a vector $x$ is in $K$ if
and only if $\alpha x \in P$ for some $\alpha > 0$. It follows
that the zero vector is also a solution to \textsc{LPcone}.
The converse is immediate since $P \subset K$.
Hence, it is sufficient to consider \textsc{LPcone}
to evaluate the performance of the LP decoder. For this reason,
we will mainly consider \textsc{LPcone} instead of the original
linear program.

\begin{figure}[ht]
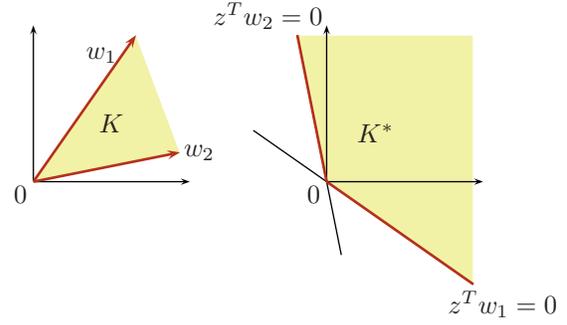

\dualcone{13mm}
\caption{The generators of the cone $K$ and the dual cone $K^*$}
\label{Fig:dualcone}
\end{figure}

To compute the probability of error, we need to find the set $K^*$
such that the zero vector is a solution to the linear program
if and only if $\gamma \in K^*$. To describe the set $K^*$,
we proceed as follows. Let $W = \{w_1,\ldots,w_m\}$
be the set of ``generators''
of the cone $K$ (Fig.~\ref{Fig:dualcone}), i.e., a vector $x$ is in $K$ 
if and only if $x$ can be written
as a nonnegative linear combination of the generators:
$x = \alpha_1 w_1 + \cdots + \alpha_m w_m$, where $\alpha_1,
\ldots , \alpha_m$ are some nonnegative real numbers.
The zero vector is a solution to the linear program
if and only if $\gamma^T x \geq 0$ for all $x \in K$.
It can be shown that this condition is equivalent to 
$\gamma^T x \geq 0$ for all $x \in W$. Hence, the decoding
is successful if and only if the log-likelihood ratio $\gamma$
is in 
\begin{equation}
\label{eq:dualcone}
K^* = \{z \in \mathbb{R}^n \; | \; z^Tx \geq 0 \; \mbox{for
all $x \in W$}\}
\end{equation}
(Fig.~\ref{Fig:dualcone}). The set $K^*$
is called the {\em{dual cone}} of $K$~\cite{BV04}.

\section{Pseudo-Weight on the AWGN Channel}
\label{Sec:pwAWGN}

For ML decoding over a memoryless channel, 
the probability of error of a code is largely determined by
its Hamming weight spectrum. For iterative and LP decoding,
it has been observed that the notion of ``pseudo-weight'' is more
appropriate for determining the probability of 
error~\cite{Wib96,FKKR01,KV03}.
The definition of pseudo-weight varies with the channel;
we will study only the pseudo-weight on the additive
white Gaussian noise (AWGN) channel.

Before stating the definition of pseudo-weight,
we will first extend the discussion in the previous section
for the AWGN channel. We hope that by doing so
the intuition behind the definition of pseudo-weight
will be more apparent.

\begin{figure}[ht]
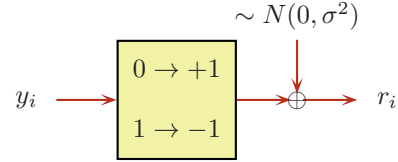

\AWGNchannel{8mm}
\caption{Binary-input AWGN channel}
\label{Fig:AWGNchannel}
\end{figure}

Consider the discrete-time AWGN channel in Fig.~\ref{Fig:AWGNchannel}.
Each bit of the codeword $y$ is modulated to $+1$ and $-1$
and then corrupted by additive white Gaussian noise with
variance $\sigma^2$. The received vector is denoted by $r$.
It can be shown that the log-likelihood ratio $\gamma$ is
given by $\gamma_i = (2/\sigma^2)r_i$. We recall from the
previous section that LP decoding is successful if and only if
$\gamma \in K^*$, the dual of the fundamental cone.
Since scaling the cost function by a positive scalar
does not change the solution
to a linear program, the decoding is successful if and only if
$r \in K^*$.

\begin{figure}[ht]
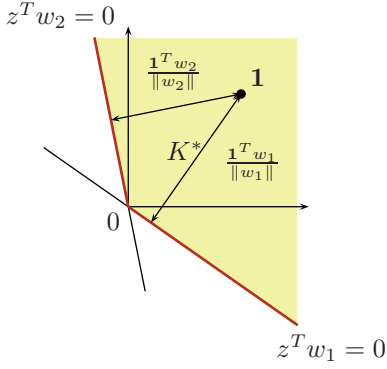

\bounddist{15mm}
\caption{Distances from the all-ones vector $\mathbf{1}$ to the hyperplanes
defining the dual cone $K^*$}
\label{Fig:distance}
\end{figure}

We recall that $W = \{w_1, \ldots, w_m\}$ 
is the set of generators for the fundamental cone.
The dual cone $K^*$ can be described by the hyperplanes $z^Tw_i = 0$
as in~(\ref{eq:dualcone}). The transmitted vector corresponding
to the all-zeros codeword is the all-ones vector, which will be
denoted by $\mathbf{1}$. The Euclidean distance from the all-ones vector
to the hyperplane $i$ is $\mathbf{1}^Tw_i / \|w_i\|$, where $\|\cdot\|$
denotes the Euclidean norm (Fig.~\ref{Fig:distance}).
If the noise perturbs the transmitted vector by this distance
in the direction perpendicular to the hyperplane,
the LP decoder will fail.

By way of comparison, consider ML decoding. In ML decoding,
the Voronoi region for the all-zeros codeword is defined by
the hyperplanes separating the all-ones vector and the transmitted
vectors of the other codewords. Let $y$ be a codeword with
Hamming weight $d$. The distance from the all-ones vector $\mathbf{1}$
to the hyperplane separating $\mathbf{1}$ and the transmitted vector 
corresponding to $y$ 
is $\sqrt{d}$. This relationship between the Euclidean distance
to the hyperplane
and the Hamming weight of $y$ motivates the definition of {\em{pseudo-weight}}
of a vector $x$ for the AWGN channel~\cite{Wib96,FKKR01,KV03}:
$$
p(x) = \left(\frac{\mathbf{1}^Tx}{\|x\|}\right)^2.
$$
If $x$ is a $0$-$1$ vector, its pseudo-weight is equal to its
Hamming weight.

Given a generator of $K$, computing its pseudo-weight is trivial.
However, given a cone, finding its generators is a very complex task.
A straightforward way is to add an equality constraint
to \textsc{LPcone}, such as $\mathbf{1}^Tx = 1$, so that the feasible set
becomes bounded. Then change the problem into the standard form
and find all ``basic feasible solutions,'' which correspond
to the corner points of the feasible region. This involves
$\binom{N}{M}$ ways of choosing ``basic variables,'' where
$N$ and $M$ are the number of variables and the number of 
equality constraints of the modified linear program. 
(For more details, refer to any linear optimization book,
e.g.,~\cite{NS96}.)

A simpler way to find the set of generators of a cone is presented
in~\cite{CCJP99}. However, the complexity is still very high
and thus the algorithm can only be applied to very short codes.
Using this algorithm, we computed the histograms of the 
pseudo-weights of the $(7,4)$ and $(15,11)$
Hamming codes, shown in Fig.~\ref{Fig:pwspectrum}.
\begin{figure}[ht]
\pwspectrum
\caption{Pseudo-weight spectra of the $(7,4)$ and the $(15,11)$
Hamming codes}
\label{Fig:pwspectrum}
\end{figure}

\section{Bounds on Minimum Pseudo-Weight}
\label{Sec:bounds}

Tanner~\cite{Tan01} gave several lower bounds on the minimum Hamming weight
of linear block codes. One of them, called the {\em{bit-oriented bound}}, is
a function of the column and row weights of the parity-check matrix $H$
and the eigenvalues of $H^TH$. Another one, called the {\em{optimization
distance bound}}, is computed by solving an optimization problem.
Two lower bounds on minimum pseudo-weight were presented in~\cite{VK04}.
One of them is similar to the bit-oriented bound of Tanner~\cite{Tan01}.
The other, called the {\em{LP-based bound}},
is computed by solving a linear program derived from 
the fundamental cone.

In this section, we will present two lower bounds on minimum pseudo-weight.
Before doing so, we prove two propositions which are useful
in establishing the bounds.

\begin{Pro} \label{Pro:minpw}
Let $K$ be a polyhedral cone and $W$ its set of generators. Then
$$
\min_{x \in W} p(x) = \min_{x \in K} p(x) = \min_{x \in K, \; \mathbf{1}^Tx = 1} p(x).
$$
\end{Pro}

The proof of Proposition~\ref{Pro:minpw} is given in the appendix.
Since $p(x) = ((\mathbf{1}^Tx)/\|x\|)^2$, it follows from Proposition~\ref{Pro:minpw}
that the problem of minimizing 
the pseudo-weight over $W$ becomes the following non-convex problem,
which we call \textsc{MaxNorm}:
\begin{center}
\begin{fminipage}
\begin{tabbing}
maximize \rule{5mm}{0mm} \= $\|x\|^2$\\
subject to \> $x \in K$,\\
	   \> $\mathbf{1}^Tx = 1$.
\end{tabbing}
\end{fminipage}
\end{center}

Our lower bounds are obtained by relaxing this difficult problem
to an easier one, particularly the one in Proposition~\ref{Pro:firstbound}
below. Since the feasible set of \textsc{MaxNorm} is contained
in the feasible set of the relaxed problem, $\|x^*\|^2 \leq \|x'\|^2$,
where $x^*$ and $x'$ are the solutions to
\textsc{MaxNorm} and the relaxed problem, respectively.
Therefore the minimum pseudo-weight, which equals $1/\|x^*\|^2$,
is lower bounded by $1/\|x'\|^2$.

\begin{Pro} \label{Pro:firstbound}
Let $\alpha_i$, $1 \leq i \leq n$, be nonnegative real numbers.
Consider the optimization problem:
\begin{center}
\begin{fminipage}
\begin{tabbing}
maximize \rule{5mm}{0mm} \= $\|x\|^2$\\
subject to \> $x \in \mathbb{R}^n$\\
           \> $0 \leq x_i \leq \alpha_i$ for all $1 \leq i \leq n$,\\
	   \> $\mathbf{1}^Tx = 1$.
\end{tabbing}
\end{fminipage}
\end{center}
Suppose that $\sum_{i=1}^{n} \alpha_i \geq 1$ and
$\alpha_i$ are ordered such that $\alpha_1 \geq \alpha_2
\geq \cdots \geq \alpha_n$. Let $j$ be the first index
such that $\alpha_1 + \cdots + \alpha_j \geq 1$. Then the
maximum of the objective function is $\alpha_1^2 + \cdots +
\alpha_{j-1}^2 + (1-\alpha_1-\cdots - \alpha_{j-1})^2$.
\end{Pro}

\begin{proof}
Let $x$ satisfy $\mathbf{1}^Tx = 1$ and $0 \leq x_i \leq \alpha_i$ 
for all $1 \leq i \leq n$.
Let $k$ be the smallest index such that $x_k < \alpha_k$.
Let $m$ be the largest index such that $x_m > 0$. We define a
new vector $x'$ as follows.

Case 1: $\alpha_k - x_k \leq x_m$. Let $x'_k = \alpha_k$
and $x'_m = x_m - \alpha_k + x_k$.

Case 2: $\alpha_k - x_k > x_m$. Let $x'_k =
x_k + x_m$ and $x'_m = 0$.

For the other indices $t \neq k, t \neq m$, let $x'_t = x_t$.
It can be shown that $\mathbf{1}^Tx' = 1$ and $0 \leq x'_i \leq \alpha_i$. 
Moreover, we claim that $\|x'\| \geq \|x\|$. 
We repeat this assignment until $x' = x$. (The algorithm terminates
since either $k$ is increased or $m$ is decreased in the next
iteration.) The final vector $x^*$ satisfies $x^*_v = \alpha_v$
for $1 \leq v \leq j-1$, $x^*_j = 1 - \alpha_1
- \cdots - \alpha_{j-1}$, and $x^*_v = 0$ for $j+1 \leq v \leq n$,
and the proposition follows.

To prove the claim, consider the two cases.

Case 1: $\alpha_k - x_k \leq x_m$.
\begin{eqnarray*}
\|x'\|^2 - \|x\|^2 & = & (x'_k)^2 + (x'_m)^2 - x_k^2 - x_m^2\\
& = & \alpha_k^2 + (x_m - \alpha_k + x_k)^2 - x_k^2 - x_m^2\\
& = & 2(\alpha_k - x_m)(\alpha_k - x_k) \geq 0.
\end{eqnarray*}

Case 2: $\alpha_k - x_k > x_m$.
$$
\|x'\|^2 - \|x\|^2 = (x_k + x_m)^2 - x_k^2 - x_m^2 \geq 0.
$$
\end{proof}

\subsection{Bound from Column Weight}

If the Tanner graph of a parity-check matrix $H$ has no cycle of length
four, it is well known that the minimum Hamming distance is 
lower bounded by the minimum column weight of $H$ plus one.
This is true for the minimum pseudo-weight as well.

\begin{Thm} \label{Thm:cwbound}
Suppose that any two columns of the parity-check matrix $H$ have 
at most one $1$
in the same position. Then the minimum pseudo-weight is lower bounded
by the minimum column weight plus one.
\end{Thm}


\begin{proof}
Let $m^*$ be the minimum column weight of $H$.
The basic idea of the proof is to relax \textsc{MaxNorm}
to the problem in Proposition~\ref{Pro:firstbound}
where $\alpha_i = 1/(m^*+1)$. Then the theorem will follow.

Consider the $i$-th column of $H$, which is denoted by $c_i$. 
Let $m$ be its column weight.
Let $q_1,\ldots,q_m$ be the positions of $1$ in $c_i$.
Since any two columns of $H$ have at most one $1$
in the same position, the support of the $q_j$-th row can be written as
$R_j \cup \{i\}$, where $R_j \cap R_k = \emptyset$ for $1 \leq k \leq m$,
$k \neq j$.

Let $x \in K$ and $\mathbf{1}^Tx = 1$.
We recall that $K$ is defined by the 
inequalities~(\ref{eq:polytopeinequality}) that
are active at the origin.
(These are the inequalities with $|V| = 1$.)
Therefore $x$ satisfies $x_i - \sum_{k \in R_j} x_k
\leq 0$ for all $1 \leq j \leq m$.
Since the sets $R_j$ are pairwise disjoint,
$$
1 = \sum_{l=1}^{n} x_l \geq x_i + \sum_{j=1}^{m} \sum_{k \in R_j} x_k
\geq x_i + \sum_{j=1}^{m} x_i = (m+1)x_i.
$$

Hence $x_i \leq 1/(m^*+1)$ for
all $1 \leq i \leq n$. Then the theorem follows from 
Proposition~\ref{Pro:firstbound}.
\end{proof}

\subsection{Relaxation Bounds}

Suppose that we choose a set $S \supseteq K$ and relax \textsc{MaxNorm}
to the following:
\begin{center}
\begin{fminipage}
\begin{tabbing}
maximize \rule{5mm}{0mm} \= $\|x\|^2$\\
subject to \> $x \in S$,\\
	   \> $\mathbf{1}^Tx = 1$.
\end{tabbing}
\end{fminipage}
\end{center}
The set $S$ should be as small as possible;
however, the new problem should be easy to solve.
A good choice for $S$ is the hyper-rectangle
$$
S = \{x \in \mathbb{R}^n \; | \; 0 \leq x_i \leq \alpha_i\},
$$
where $\alpha_i$ is the maximum of the objective function
of the following linear program:
\begin{center}
\begin{fminipage}
\begin{tabbing}
maximize \rule{5mm}{0mm} \= $x_i$\\
subject to \> $x \in K$,\\
	   \> $\mathbf{1}^Tx = 1$.
\end{tabbing}
\end{fminipage}
\end{center}
Without loss of generality, assume that $\alpha_1 \geq \alpha_2 \geq
\cdots \geq \alpha_n$. Let $j$ be the first index such that
$\alpha_1 + \cdots + \alpha_j \geq 1$. From Proposition~\ref{Pro:firstbound},
$$
\mbox{min pseudo-weight} \geq \frac{1}{\sum_{i=1}^{j-1} \alpha_i^2
+ (1-\sum_{i=1}^{j-1} \alpha_i)^2}
$$
We call this the \textit{first-order bound}.

A more complex choice for $S$ is
$$
S = \{x \in \mathbb{R}^n \; | \; 0 \leq x_i \leq \alpha_i, \;
x_i + x_j \leq \beta_{i,j}\},
$$
where $\alpha_i$ are computed in the same way
as above, and $\beta_{i,j}$ are computed by replacing the
objective function of the above linear program by $x_i + x_j$.
Unfortunately, the problem is hard to solve for this set $S$.
We approach the problem by partitioning the feasible region into
$n$ sub-regions based on the maximum entry of $x$.
We obtain $n$ sub-problems; the sub-problem $k$, $1 \leq k \leq n$,
is the following:
\begin{center}
\begin{fminipage}
\begin{tabbing}
maximize \rule{5mm}{0mm} \= $\|x\|^2$\\
subject to \> $0 \leq x_i \leq \alpha_i$ \rule{5mm}{0mm} \= for all $1 \leq i \leq n$,\\
           \> $x_i + x_j \leq \beta_{i,j}$ \> for all $1 \leq i \leq n$,\\
	   \> \> $1 \leq j \leq n$,\\
	   \> $\mathbf{1}^Tx = 1$,\\
	   \> $x_k \geq x_i$ \> for all $1 \leq i \leq n$.
\end{tabbing}
\end{fminipage}
\end{center}

Then we relax each sub-problem by omitting the inequalities
$x_i + x_j \leq \beta_{i,j}$ whenever $i \neq k$ and $j \neq k$.
\begin{center}
\begin{fminipage}
\begin{tabbing}
maximize \rule{5mm}{0mm} \= $\|x\|^2$\\
subject to \> $0 \leq x_i \leq \alpha_i$ \rule{5mm}{0mm} \= for all $1 \leq i \leq n$,\\
           \> $x_k + x_i \leq \beta_{k,i}$ \> for all $1 \leq i \leq n$,\\
	   \> $\mathbf{1}^Tx = 1$,\\
	   \> $x_k \geq x_i$ \> for all $1 \leq i \leq n$.
\end{tabbing}
\end{fminipage}
\end{center}
If we fix $x_k$, then the relaxed sub-problem has the same form as when $S$
is a rectangle, which we can solve by Proposition~\ref{Pro:firstbound}. 
Thus we can compute the maximum of the relaxed sub-problem for
a fixed $x_k$. By varying $x_k$ and computing the corresponding
maximum, we can solve the relaxed sub-problem $k$. 
Taking the maximum of $\|x\|^2$
over all relaxed sub-problems, we obtain an upper bound of $\|x\|^2$
for the original problem, which will lead to a lower bound for
the minimum pseudo-weight.
We call this the \textit{second-order bound}.







We have a few remarks regarding these lower bounds:
\begin{itemize}
\item Any lower bound for minimum pseudo-weight is also a lower bound
for minimum Hamming weight.

\item To compute the
first-order and the second-order bounds, we need to solve
as many as $n$ and $n(n+1)/2$ linear programs respectively.
Each of these linear programs has $n$ variables. This is in contrast
to the optimization distance bound and the LP-based bound discussed
earlier, which are computed by solving one linear program whose
worst-case number of variables is quadratic in the codeword length.

\item The number of linear programs to be solved for
the first-order and the second-order bounds can be reduced if the code
has some structure.
\end{itemize}

Next we compute these bounds for
a class of group-structured LDPC codes presented in~\cite{TSF01}.
A particular sequence of codes in this class has constant column
weight 3 and constant row weight 5, with rate approximately $2/5$.
The bounds on minimum pseudo-weights for these codes are shown in
Table~\ref{Tbl:ldpcbounds}.
The upper bound is computed by using the MATLAB optimization
toolbox to find a good feasible solution to \textsc{MaxNorm}.
Note that for the code with length $155$, Vontobel and Koetter~\cite{VK04}
have found tighter lower and upper bounds: $10.8$ and $16.4$ respectively.

\begin{table}[ht]
\caption{Lower and upper bounds on minimum pseudo-weights
of Tanner's LDPC codes with rate approximately $2/5$}
\label{Tbl:ldpcbounds}
\begin{center}
\begin{tabular}{|c|c|c|c|} \hline
length & first-order & second-order & upper bound\\ \hline
155 & 8.3 & 9.7 & 17.0\\
305 & 11.5 & 13.8 & 20.1\\
755 & 13.0 & 14.0 & 27.6\\
905 & 17.6 & 21.5 & 39.4\\ \hline
\end{tabular}
\end{center}
\end{table}

\appendix[Proof of Proposition~\ref{Pro:minpw}]

In this appendix we present a proof of Proposition~\ref{Pro:minpw}.
First, we need the following lemma.
\begin{Lem} \label{Lem:fraction}
Let $a,b,c,d$ be positive real numbers. Then
$$
\frac{a+c}{b+d} \geq \min\left\{\frac{a}{b}, \frac{c}{d}\right\}.
$$
\end{Lem}

\begin{proof}
Suppose that $a/b \leq c/d$. Then
\begin{eqnarray*}
\frac{c}{a} & \geq & \frac{d}{b}\\
\frac{a+c}{a} & \geq & \frac{b+d}{b}\\
\frac{a+c}{b+d} & \geq & \frac{a}{b}.
\end{eqnarray*}
The case $c/d \leq a/b$ is similar.
\end{proof}

\begin{proofof}{Proposition~\ref{Pro:minpw}}
First, we will show that $\min_{x \in W} p(x) = \min_{x \in K} p(x)$.
Since $W \subset K$, $\min_{x \in W} p(x) \geq \min_{x \in K} p(x)$.
Conversely, let $x,y,z \in K$ and $\alpha, \beta > 0$ 
with $x = \alpha y + \beta z$.
Then $\mathbf{1}^Tx = \alpha \mathbf{1}^Ty + \beta \mathbf{1}^Tz$. 
By the triangle inequality, $\|x\| \leq \alpha \|y\| + \beta \|z\|.$
It follows that
\begin{eqnarray*}
p(x) &=& \left(\frac{\mathbf{1}^Tx}{\|x\|}\right)^2
\geq \left(\frac{\alpha \mathbf{1}^Ty + \beta \mathbf{1}^Tz}{\alpha \|y\|+\beta \|z\|}\right)^2\\
&\geq& \min \left\{\left(\frac{\mathbf{1}^Ty}{\|y\|}\right)^2,
\left(\frac{\mathbf{1}^Tz}{\|z\|}\right)^2\right\} = \min \{p(y), p(z)\},
\end{eqnarray*}
where the latter inequality follows from Lemma~\ref{Lem:fraction}.
Since every $x \in K$ can be written as a nonnegative linear combination
of the generators, there is a generator $w \in W$ such that
$p(x) \geq p(w)$. Therefore $\min_{x \in W} p(x) \leq \min_{x \in K} p(x)$.
Finally, $\min_{x \in K} p(x) = \min_{x \in K, \; \mathbf{1}^Tx = 1} p(x)$
since the pseudo-weight is invariant under scaling, i.e., $p(ax) = p(x)$.
\end{proofof}

\section*{Acknowledgement}
The authors would like to thank the anonymous reviewers for many
helpful comments and suggestions.

This work is supported in part by the Center for Magnetic Recording Research,
the Information Storage Industry Consortium, and the NSF under Grant
CCR-0219852.

\bibliographystyle{IEEEtranS}
\bibliography{IEEEabrv,constraint}

\end{document}